\documentclass[10pt]{article}

\usepackage[a4paper,margin=2.5cm]{geometry}
\def\md{d\kern-0.035cm\char39\kern-0.03cm}
\def\ml{l\kern-0.035cm\char39\kern-0.03cm}
\def\mt{t\kern-0.035cm\char39\kern-0.03cm}
\def\mL{L\kern-0.08cm\char39}

\usepackage{tikz}
\usetikzlibrary{arrows.meta,positioning}

\usepackage{amssymb,amsmath}
\usepackage{verbatim} 
\usepackage{amsthm}
\usepackage{color} 
\usepackage{url}
\usepackage{lineno}

\def\s2{\sigma^2}

\usepackage{hyperref}
\usepackage[numbers,sort&compress]{natbib}
\usepackage{caption}
\usepackage{float}

\usepackage{xcolor}
\definecolor{darkgreen}{rgb}{0.0, 0.6, 0.0}
\definecolor{darkred}{rgb}{0.85, 0.0, 0.0}

\usepackage{tabularx}
\usepackage{array}

\usepackage{mathtools} 
\usepackage{graphicx}  
\usepackage{dcolumn}   
\usepackage{bm}        

\begin{document}

\title{Observation delays can bias inference of selective advantage in evolutionary competition}
\author{
Robert Vala{\v s}ka,
Katar{\'\i}na Bo{\md}ov\'a\thanks{katarina.bodova@fmph.uniba.sk} \thanks{ORCID: 0000-0002-7214-0171}
\\
\small Faculty of Mathematics, Physics, and Informatics\\
\small Comenius University, Mlynsk{\' a} Dolina, 84248, Bratislava, Slovakia
}


\date{}
\maketitle
\begin{abstract}
Relative-frequency trajectories are often used to infer selective advantage in competing biological populations. A common empirical approach is to fit a linear function to the logit-transformed frequency of an invading type and interpret the slope as the relative advantage. Here we test how this estimator is affected when the competing types are observed after type-specific delays. We use SARS-CoV-2 variant replacement in the United Kingdom as empirical motivation and study the mechanism with simple two-type models. In an ideal exponential replacement model, fixed or randomly distributed observation delays change the intercept of the observed log-odds trajectory but not its slope, provided that delayed counts are aggregated before frequencies are formed and boundary effects are absent. In nonlinear SIR-type models, where the relative growth is time-dependent, delays can compare the competing variants at different dynamical phases and substantially bias the fitted logit slope. This bias can occur even when the delayed replacement trajectory remains nearly linear on the logit scale. Thus, a good logistic fit to the observed data is not sufficient to guarantee that the inferred logit slope approximates well the relative growth advantage. Observation delays should therefore be accounted for when logit slopes are used as proxies for relative fitness in nonlinear evolutionary competition.
\end{abstract}

\section{Introduction}

Evolutionary competition between coexisting types is commonly studied through changes in their relative frequencies. This perspective applies across a wide range of biological systems, from genetic and microbial populations to viral and cellular competition. In evolutionary dynamics \citep{HofbauerSigmund1998,HofbauerSigmund2003,Nowak2006}, frequency changes are driven by differences in growth rates or fitness values. The same principle is used experimentally, for example to estimate relative fitness from viral competition assays \citep{Maree2000}. When the relative growth difference is approximately constant, the frequency of the invading type follows an approximately logistic trajectory, so the slope of the logit trajectory provides a phenomenological estimate of selective advantage. Beyond this simple logit-slope approximation, selection coefficients have also been inferred from time-series allele-frequency data using likelihood-based methods \citep{Bollback2008,Malaspinas2012,Feder2014}, although linkage, clonal interference, and genetic hitchhiking can complicate such inference in asexual or rapidly adapting populations \citep{IllingworthMustonen2011}. Changes in genotype or lineage frequencies are likewise used to quantify adaptation and competitive fitness in long-term microbial evolution experiments \citep{Lenski1991,Barrick2009,Good2017}.

Observed frequencies are usually not direct observations of the underlying process. Events are typically recorded after sampling and further processing. If the relevant delays are common to both competing types, they may mainly shift the replacement trajectory in time. On the other hand, different delays result in a comparison of the two types at different phases and consequently in a change in the fitted logit slope. The main question considered here is how reliable is logit-slope inference of relative advantage when the observed frequency trajectory is delayed relative to the underlying dynamics.

SARS-CoV-2 variant replacement provides a concrete example of this problem. During the COVID-19 pandemic, SARS-CoV-2 repeatedly evolved lineages with increased transmission potential, leading to rapid replacement events at large geographic scales \citep{Davies2021,Volz2021,McCrone2022,Elliott2022,Viana2022}. Replacement events were observed through genomic surveillance, and sufficiently dense sequence data made it possible to estimate variant growth advantages from changes in lineage frequencies \citep{Volz2021,vanDorp2021,McCrone2022,Abousamra2024,Otto2021,Otto2024}. The data are available through GISAID, along with metadata, including collection date, location, and lineage assignment \citep{ShuMcCauley2017,Khare2021}. At the same time, surveillance data depend on sampling intensity, but also on sampling characteristics such as representativeness, which can affect estimates based on observed variant frequencies \citep{Chen2022,Brito2022}. In genomic surveillance, the sample collection date occurs after infection, with delays introduced by symptom onset, testing behavior, and clinical or surveillance sampling. Additional delays due to sequencing, submission, and curation affect when a record becomes available for analysis, but they do not change the collection date assigned to the sample. Since the empirical analysis in this paper uses collection date as the observation date, the most relevant delay is the delay from infection to sample collection.

In this work, we use two major replacement events in the United Kingdom, Alpha$\to$Delta and Delta$\to$Omicron BA.1, as empirical examples. These replacement events have been widely used to summarize variant growth advantages mainly because they are well approximated by linear trends in log-odds. Therefore, it is particularly important to know whether delayed observation affects inference of relative growth advantage for these data.

Existing delay-modeling methods in epidemiology mainly address incidence nowcasting \citep{McGough2020}, correction for right truncation \citep{Seaman2022}, and estimation of epidemiological delay distributions \citep{Charniga2024}. The problem considered here is related but distinct. We do not attempt to reconstruct true infection incidence from incomplete real-time data. Instead, we ask how type-specific observation delays can affect logit-slope inference when relative frequencies are computed from delayed observations.

We first analyze an exponential replacement model, where observation delays change the log-odds intercept but not the slope under appropriate conditions. We then study two-variant SIR-type models, starting from the classical SIR framework \citep{Kermack1927,Weiss2013} and considering simple extensions with waning immunity and altered cross-immunity structure \citep{Kucharski2016}. In these nonlinear models, depletion of susceptibles and interaction between variants make the relative growth rate time-dependent. Variant-specific delays can then bias the logit-slope estimate even when the observed replacement curve remains nearly logistic. We do not aim to provide a calibrated SARS-CoV-2 forecasting model or to derive general bounds on the estimation error, but instead to identify when logit-slope inference is robust or when it may fail. Moreover, we show that two types of failure can occur: either the estimate is biased despite an apparently good logistic fit, or the estimate is not interpretable because the replacement trajectory is not approximately logit-linear.

\section{Empirical example: epidemiological replacement waves}
\label{sec:empirical}

We use SARS-CoV-2 variant replacement in the United Kingdom as a concrete example of logit-slope inference from observed frequency data. Sequence metadata were obtained from GISAID for the period from January 2020 to December 2024. Each record was treated as one sequence observation. The analysis used the accession identifier, collection date, location, and Pango lineage. Locations were aggregated to the United Kingdom as a whole and, where applicable, to England, Wales, Scotland, and Northern Ireland. Throughout the empirical analysis, the collection date is used as the observation date. We therefore do not model delays between collection, sequencing, submission, and data curation. These delays are relevant for real-time data availability but are not part of the observation-time variable analyzed here.

The empirical analysis is used to demonstrate that major SARS-CoV-2 replacement waves can be approximately logistic on the scale of observed sequence frequencies indexed by collection date. The empirical data motivate timing regimes for the numerical SIR experiments. We do not infer the real infection-to-collection delay distribution from GISAID metadata.

Let $y_1(t)$ and $y_2(t)$ denote the numbers of observed sequences assigned to a resident variant~1 and an invading variant~2 on observation date $t$. The observed sequence frequency of the invading variant is
\begin{equation}
    p_2^{\mathrm{seq}}(t)
    =
    \frac{y_2(t)}{y_1(t)+y_2(t)}.
\end{equation}
For each replacement wave, we fitted a linear model to the logit-transformed
sequence frequency,
\begin{equation}
    \operatorname{logit}p_2^{\mathrm{seq}}(t)
    =
    \log\left(\frac{p_2^{\mathrm{seq}}(t)}
    {1-p_2^{\mathrm{seq}}(t)}\right)
    =
    \alpha + s_{\mathrm{seq}} t.
    \label{eq:empirical_logit_fit}
\end{equation}
The slope $s_{\mathrm{seq}}$ is interpreted as an apparent relative growth
advantage on the sequence-observation time scale. In a pure exponential growth
regime it equals the difference between the exponential growth rates of the two
variants. In more general epidemic dynamics it is a descriptive summary of the
observed replacement trajectory and should not be read directly as an intrinsic
transmissibility parameter.

To reduce boundary effects, the fit used only time points with
$0.05<p_2(t)<0.95$. For descriptive uncertainty in the empirical figures, we
used a non-parametric bootstrap of the transformed time points with 1000
resamples. This bootstrap is used only as a robustness check for the simple
logit-slope estimator; it is not a full observation model for sequence counts.

For the main empirical analysis we used the Alpha$\to$Delta and
Delta$\to$Omicron BA.1 replacement waves. Later Omicron sublineage replacements
are described in Supplementary Material, but they are not used for the
main conclusions because their category definitions depend more strongly on
aggregation choices.

\begin{figure}[ht]
\centering
\includegraphics[width=0.98\textwidth]{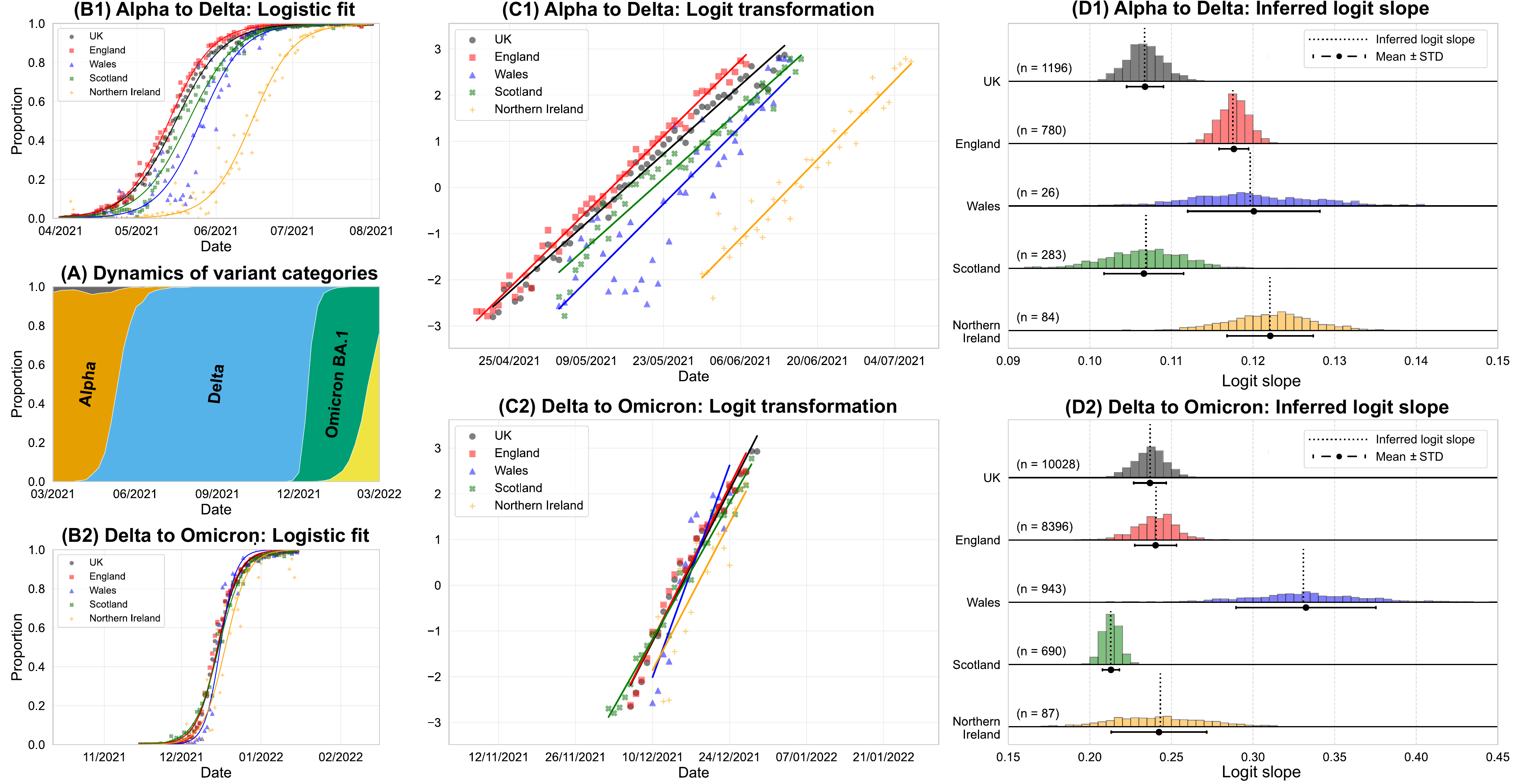}
\caption{Empirical replacement dynamics for Alpha$\to$Delta and
Delta$\to$Omicron BA.1.
{(A)} Overview of the observed sequence dynamics of Alpha, Delta, and Omicron
BA.1. In the figure label, Omicron BA.1 is shortened to Omicron.
{(B1)} Observed Delta proportions for the United Kingdom and its four regions
during the Alpha$\to$Delta replacement wave, together with fitted logistic
curves (with matching colors).
{(B2)} Analogous observed Omicron BA.1 proportions for the
Delta$\to$Omicron BA.1 replacement wave.
{(C1)} Logit transformation of the Alpha$\to$Delta replacement wave.
{(C2)} Logit transformation of the Delta$\to$Omicron BA.1 replacement wave.
{(D1)} Logit-slope estimates for Alpha$\to$Delta with descriptive uncertainty
from a non-parametric bootstrap. The reported value $n$ denotes the number of
sequence observations near the midpoint of the fitted logistic curve.
{(D2)} Analogous estimates for Delta$\to$Omicron BA.1.}
\label{fig:empirical}
\end{figure}

The Alpha$\to$Delta and Delta$\to$Omicron BA.1 replacement waves were
approximately logistic in the United Kingdom and in England, Wales, Scotland,
and Northern Ireland (Figure~\ref{fig:empirical}B). The central part of each
wave was approximately linear after logit transformation
(Figure~\ref{fig:empirical}C), supporting the use of the logit slope as a
descriptive summary of the observed replacement.

For Alpha$\to$Delta, the UK-wide logit-slope estimate was
$0.107$ day$^{-1}$, see Table~\ref{tab:empirical}. Regional estimates ranged from
$0.107$ day$^{-1}$ in Scotland to $0.122$ day$^{-1}$ in Northern Ireland.
For Delta$\to$Omicron BA.1, the UK-wide estimate was larger,
$0.237$ day$^{-1}$. These values describe observed sequence-frequency
dynamics on the observation-time scale.

\begin{table}[ht]
\centering
\caption{Logit-slope estimates, in day$^{-1}$, for two major UK SARS-CoV-2
replacement waves. The estimates summarize observed sequence-frequency
dynamics and should not be interpreted as direct transmissibility estimates.}
\label{tab:empirical}
\begin{tabular}{lccccc}
\hline
Wave & UK & England & Wales & Scotland & N. Ireland \\
\hline \hline
Alpha$\to$Delta
& 0.107 & 0.118 & 0.120 & 0.107 & 0.122 \\
Delta$\to$Omicron BA.1
& 0.237 & 0.241 & 0.331 & 0.213 & 0.243 \\
\hline
\end{tabular}
\end{table}

\section{Mathematical models}
\label{sec:models}

\subsection{Two competing populations and logit-slope inference}
\label{sec:two_type_model}

We first formulate the inference problem for two competing types. Let
$\lambda_1(t)$ and $\lambda_2(t)$ denote the latent event-time intensities of
the two types. In the epidemiological example used below, these intensities are
daily incidences of infections caused by two pathogen variants. The latent
frequency of type~2 is
\begin{equation}
    p_2(t)
    =
    \frac{\lambda_2(t)}{\lambda_1(t)+\lambda_2(t)}.
\end{equation}
The corresponding log odds are
\begin{equation}
    L(t)
    =
    \log\left(\frac{p_2(t)}{1-p_2(t)}\right)
    =
    \log\left(\frac{\lambda_2(t)}{\lambda_1(t)}\right).
\end{equation}
The instantaneous relative growth rate is
\begin{equation}
    s(t)
    =
    \frac{dL(t)}{dt}
    =
    \frac{d}{dt}\log \lambda_2(t)
    -
    \frac{d}{dt}\log \lambda_1(t).
    \label{eq:instantaneous_relative_growth}
\end{equation}
This quantity is the continuous-time analogue of a selection coefficient or
relative fitness difference. If $s(t)$ is constant, then $L(t)$ is linear in
time and $p_2(t)$ follows an exact logistic curve. If $s(t)$ varies in time, a
single fitted logit slope is a window-specific summary of the replacement
trajectory rather than a fixed structural parameter.

\subsection{Exponential replacement model}
\label{sec:exponential_model}

The simplest reference case is exponential replacement. Suppose that the two
latent intensities satisfy
\begin{equation}
    \lambda_1(t)=C_1e^{r_1t},
    \qquad
    \lambda_2(t)=C_2e^{r_2t}.
\end{equation}
Then
\begin{equation}
    L(t)
    =
    \log\left(\frac{C_2}{C_1}\right)
    +
    (r_2-r_1)t.
\end{equation}
Thus the logit slope is exactly
\begin{equation}
    s_0=r_2-r_1.
\end{equation}

\subsection{Observation model with delays}
\label{sec:delay_model}

Observed data are modeled as delayed versions of the latent event-time
intensities. Let $g_i(d)$ be the delay distribution of type $i$, where $d\geq0$
is measured in days, and let $\rho_i$ be a constant observation multiplier. The
expected observed count or intensity of type $i$ at observation time $t$ is
\begin{equation}
    y_i(t)
    =
    \rho_i
    \sum_{d\geq 0}
    \lambda_i(t-d)g_i(d).
    \label{eq:discrete_delay_model}
\end{equation}
In continuous time, the analogous expression is
\begin{equation}
    y_i(t)
    =
    \rho_i
    \int_0^\infty
    \lambda_i(t-u)g_i(u)\,du.
\end{equation}
The observed type-2 frequency is
\begin{equation}
    p_2^{\mathrm{obs}}(t)
    =
    \frac{y_2(t)}{y_1(t)+y_2(t)}.
\end{equation}
The logit-slope estimator applied to observed data is therefore based on
\begin{equation}
    L_{\mathrm{obs}}(t)
    =
    \log\left(\frac{y_2(t)}{y_1(t)}\right).
\end{equation}
For a fixed delay $d_i$, the distribution is a point mass and
\begin{equation}
    y_i(t)
    =
    \rho_i \lambda_i(t-d_i).
\end{equation}
For a non-degenerate delay distribution, the observed data are obtained by
convolving the latent intensity with the delay distribution. The
delay is part of the observation model, it does not change the underlying
population dynamics.

\subsection{Delay invariance in the exponential model}
\label{sec:exponential_delay_result}

For exponential replacement, the observation-delay model has a simple
consequence, provided that the exponential form is valid over the full range of
event times contributing to the delayed observation. Equivalently, the argument
applies away from introduction, truncation, or finite-window boundary effects.
Suppose that the latent incidence of variant $i$ is
$\lambda_i(t)=C_i e^{r_i t}$, and that observations of variant $i$ are delayed
by a non-negative random delay $D_i$ with probability mass function $g_i(d)$.
The expected observed incidence is then
\begin{equation}
    y_i(t)
    =
    \rho_i
    \sum_{d\geq 0}
    \lambda_i(t-d)g_i(d).
\end{equation}
Substituting the exponential form gives
\begin{align}
    y_i(t)&=
    \rho_i C_i
    \sum_{d\geq0}
    e^{r_i(t-d)}g_i(d)
	=
    \rho_i C_i e^{r_i t}
    \sum_{d\geq0}
    e^{-r_i d}g_i(d)
	=
    \widetilde C_i e^{r_i t},
\end{align}
where
\begin{equation}
    \widetilde C_i
    =
    \rho_i C_i
    \sum_{d\geq0} e^{-r_i d}g_i(d),
\end{equation}
provided that the corresponding transform of the delay distribution is finite.
Thus the delay distribution changes only the multiplicative constant
$\widetilde C_i$, not the exponential growth rate $r_i$. This includes the case
of a fixed delay as a special case, where $g_i$ is a point mass at $d_i$ and the
multiplicative factor becomes $e^{-r_i d_i}$.

Consequently, the observed log-odds satisfy
\begin{equation}
    L_{\mathrm{obs}}(t)
    =
    \log\left(\frac{y_2(t)}{y_1(t)}\right)
    =
    \log\left(\frac{\widetilde C_2}{\widetilde C_1}\right)
    +
    (r_2-r_1)t.
\end{equation}
Therefore, in the ideal exponential model, fixed or randomly distributed
observation delays affect the intercept of the observed log-odds trajectory but
not its slope. The logit-slope estimate remains equal to the infection-time
growth-rate difference $r_2-r_1$, as long as the exponential trajectories are not
distorted by time-dependent growth rates, as in nonlinear epidemic dynamics, discretization, or finite-window boundary effects.

\subsection{Logit slope bias for time-dependent growth}
\label{sec:general_delay_bias}

The invariance of the logit slope in the exponential model follows from the
constant nature of the growth rates. For time-dependent latent growth rates, 
the effect of fixed observation delays can be expressed exactly. 
Define the time-dependent growth rates
\begin{equation}
    r_i(t)
    =
    \frac{d}{dt}\log \lambda_i(t),
\end{equation}
so that the instantaneous relative growth rate introduced in
Eq.~\eqref{eq:instantaneous_relative_growth} is
\begin{equation}
    s(t)=r_2(t)-r_1(t).
\end{equation}
Consider fixed type-specific observation delays $d_1$ and $d_2$. From
\begin{equation}
    y_i(t)=\rho_i\lambda_i(t-d_i),
\end{equation}
the observed log odds are
\begin{equation}
    L_{\mathrm{obs}}(t)
    =
    \log\frac{\rho_2}{\rho_1}
    +
    \log\lambda_2(t-d_2)
    -
    \log\lambda_1(t-d_1).
\end{equation}
Differentiating with respect to observation time gives the exact relation
\begin{equation}
    s_{\mathrm{obs}}(t)
    =
    \frac{dL_{\mathrm{obs}}(t)}{dt}
    =
    r_2(t-d_2)-r_1(t-d_1).
    \label{eq:fixed_delay_slope}
\end{equation}
Consequently, the difference between the observed-time and event-time
instantaneous relative growth rates is
\begin{align}
    s_{\mathrm{obs}}(t)-s(t)
    &=
    \bigl[r_2(t-d_2)-r_2(t)\bigr]
    -
    \bigl[r_1(t-d_1)-r_1(t)\bigr].
    \label{eq:fixed_delay_bias_exact}
\end{align}
This expression shows that a delay changes the inferred relative
growth only when the individual growth rates vary with time. In the
exponential case, $r_i(t)=r_i$ is constant and
Eq.~\eqref{eq:fixed_delay_bias_exact} vanishes exactly, recovering the
invariance result above.

If both types have the same delay, $d_1=d_2=d$, then
Eq.~\eqref{eq:fixed_delay_slope} reduces to
\begin{equation}
    s_{\mathrm{obs}}(t)=s(t-d).
\end{equation}
A common delay therefore produces only a time shift of the instantaneous
relative-growth trajectory. In contrast, when $d_1\neq d_2$, the two growth
rates entering Eq.~\eqref{eq:fixed_delay_slope} are evaluated at different
dynamical times. This asynchronous comparison is the mechanism by which
type-specific observation delays can alter the apparent selective
advantage in nonlinear dynamics.

\subsection{Two-variant SIR models}
\label{sec:SIRmodels}

\begin{figure}
\centering
\begin{tikzpicture}[>=Stealth,
    compartment/.style={draw,rounded corners,minimum width=0.8cm,minimum height=0.8cm,align=center},
    lbl/.style={font=\small}]
\node[font=\bfseries] at (3,2.7) {A};
\node[compartment] (S) at (0,0) {$S$};
\node[compartment] (IA) at (3,1.8) {$I_1$};
\node[compartment] (IB) at (3,-1.8) {$I_2$};
\node[compartment] (R) at (6,0) {$R$};
\draw[->] (S) -- node[lbl, above left] {$\dfrac{\beta_1}{N}I_1$} (IA);
\draw[->] (S) -- node[lbl, below left] {$\dfrac{\beta_2}{N}I_2$} (IB);
\draw[->] (IA) -- node[lbl, above] {$\gamma_1$} (R);
\draw[->] (IB) -- node[lbl, below] {$\gamma_2$} (R);
\end{tikzpicture}
\hskip 0.3cm
\begin{tikzpicture}[>=Stealth,
    compartment/.style={draw,rounded corners,minimum width=0.8cm,minimum height=0.8cm,align=center},
    lbl/.style={font=\small}]
\node[font=\bfseries] at (12,2.7) {B};
\node[compartment] (S) at (9,0) {$S$};
\node[compartment] (IA) at (12,1.8) {$I_1$};
\node[compartment] (IB) at (12,-1.8) {$I_2$};
\node[compartment] (R) at (15,0) {$R$};
\draw[->] (S) -- node[lbl, above left] {$\dfrac{\beta_1}{N}I_1$} (IA);
\draw[->] (S) -- node[lbl, below left] {$\dfrac{\beta_2}{N}I_2$} (IB);
\draw[->] (IA) -- node[lbl, above] {$\gamma_1$} (R);
\draw[->] (IB) -- node[lbl, below] {$\gamma_2$} (R);
\draw[->] (R) -- node[lbl, above right] {$\omega$} (S);
\end{tikzpicture}

\vskip 0.3cm

\begin{tikzpicture}[>=Stealth,
    compartment/.style={draw,rounded corners,minimum width=0.8cm,minimum height=0.8cm,align=center},
    lbl/.style={font=\small}]
\node[font=\bfseries] at (6,2.7) {C};
\node[compartment] (S) at (0,0) {$S$};
\node[compartment] (IA) at (2,1.8) {$I_1$};
\node[compartment] (RA) at (6,1.8) {$R_1$};
\node[compartment] (IBA) at (10,1.8) {$I_{21}$};
\node[compartment] (IB) at (2,-1.8) {$I_2$};
\node[compartment] (RB) at (6,-1.8) {$R_2$};
\node[compartment] (IAB) at (10,-1.8) {$I_{12}$};
\node[compartment] (RAB) at (12,0) {$R_{12}$};
\draw[->] (S) -- node[lbl, above left] {$\dfrac{\beta_1}{N}\left(I_1+I_{12}\right)$} (IA);
\draw[->] (S) -- node[lbl, below left] {$\dfrac{\beta_2}{N}\left(I_2+I_{21}\right)$} (IB);
\draw[->] (IA) -- node[lbl, above] {$\gamma_1$} (RA);
\draw[->] (IB) -- node[lbl, below] {$\gamma_2$} (RB);
\draw[->] (RA) -- node[lbl, above] {$\dfrac{\beta_2}{N}\left(I_2+I_{21}\right)$} (IBA);
\draw[->] (RB) -- node[lbl, below] {$\dfrac{\beta_1}{N}\left(I_1+I_{12}\right)$} (IAB);
\draw[->] (IBA) -- node[lbl, above right] {$\gamma_2$} (RAB);
\draw[->] (IAB) -- node[lbl, below right] {$\gamma_1$} (RAB);
\end{tikzpicture}

\vskip 0.3cm

\begin{tikzpicture}[>=Stealth,
    compartment/.style={draw,rounded corners,minimum width=0.8cm,minimum height=0.8cm,align=center},
    lbl/.style={font=\small}]
\node[font=\bfseries] at (6,2.7) {D};
\node[compartment] (S) at (0,0) {$S$};
\node[compartment] (IA) at (2,1.8) {$I_1$};
\node[compartment] (RA) at (6,1.8) {$R_1$};
\node[compartment] (IBA) at (10,1.8) {$I_{21}$};
\node[compartment] (IB) at (2,-1.8) {$I_2$};
\node[compartment] (RB) at (6,-1.8) {$R_2$};
\node[compartment] (IAB) at (10,-1.8) {$I_{12}$};
\node[compartment] (RAB) at (12,0) {$R_{12}$};
\draw[->] (S) -- node[lbl, above left] {$\dfrac{\beta_1}{N}\left(I_1+I_{12}\right)$} (IA);
\draw[->] (S) -- node[lbl, below left] {$\dfrac{\beta_2}{N}\left(I_2+I_{21}\right)$} (IB);
\draw[->] (IA) -- node[lbl, above] {$\gamma_1$} (RA);
\draw[->] (IB) -- node[lbl, below] {$\gamma_2$} (RB);
\draw[->] (RA) -- node[lbl, above] {$\chi_{21}\dfrac{\beta_2}{N}\left(I_2+I_{21}\right)$} (IBA);
\draw[->] (RB) -- node[lbl, below] {$\chi_{12}\dfrac{\beta_1}{N}\left(I_1+I_{12}\right)$} (IAB);
\draw[->] (IBA) -- node[lbl, above right] {$\gamma_2$} (RAB);
\draw[->] (IAB) -- node[lbl, below right] {$\gamma_1$} (RAB);
\draw[->] (RA) -- node[lbl, above right] {$\omega_1$} (S);
\draw[->] (RB) -- node[lbl, above right] {$\omega_2$} (S);
\draw[->] (RAB) -- node[lbl, above right] {$\omega_2$} (RA);
\draw[->] (RAB) -- node[lbl, above right] {$\omega_1$} (RB);
\end{tikzpicture}

\caption{Compartment diagrams for the SIR-type models used in the numerical
experiments. 
A. Baseline model with complete cross-immunity.
B. Complete cross-immunity with waning immunity.
C. No cross-immunity.
D. Combined model with partial asymmetric protection and variant-specific
waning immunity. The arrow labels are per-capita transition rates and must be
multiplied by the population size of the source compartment. In the double-index
notation, $I_{12}$ denotes individuals previously infected by variant~2 and
currently infected by variant~1, while $I_{21}$ denotes individuals previously
infected by variant~1 and currently infected by variant~2. The recovered
classes $R_{12}$ and $R_{21}$ are identical.}
\label{fig:sir_diagrams}
\end{figure}
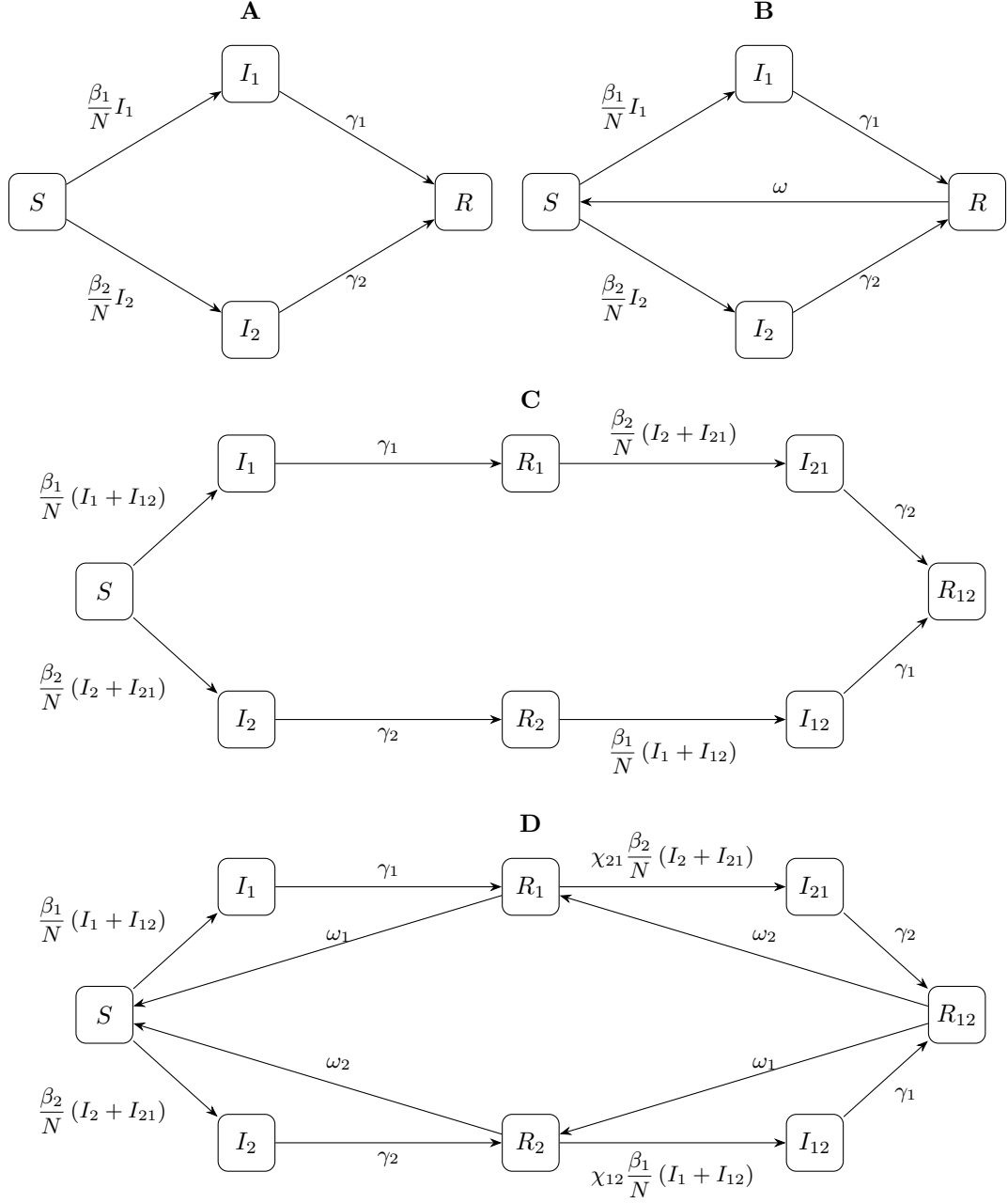

We next use two-variant SIR-type models \cite{Kermack1927} in which the
relative growth rate changes over time. The baseline model, shown in Figure~\ref{fig:sir_diagrams} (A) has complete
cross-immunity:
\begin{align}
    \frac{dS}{dt}
    &=
    -\frac{\beta_1}{N}SI_1
    -\frac{\beta_2}{N}SI_2,
    \\
    \frac{dI_1}{dt}
    &=
    \frac{\beta_1}{N}SI_1-\gamma_1I_1,
    \\
    \frac{dI_2}{dt}
    &=
    \frac{\beta_2}{N}SI_2-\gamma_2I_2,
    \\
    \frac{dR}{dt}
    &=
    \gamma_1I_1+\gamma_2I_2.
\end{align}
Variant~1 is introduced first and variant~2 is introduced later. The daily
incidence of variant $i$ is computed from the infection term,
\begin{equation}
    \lambda_i(t)
    =
    \frac{\beta_i}{N}S(t)I_i(t).
\end{equation}
The delay model in Eq.~\eqref{eq:discrete_delay_model} is then applied to
$\lambda_i(t)$, not to prevalence $I_i(t)$.

We also considered three extensions of the baseline model to test whether the delay effect persists under changes in immunity structure:
\begin{enumerate}
    \item Model with complete cross-immunity and waning immunity, in which
    recovered individuals return to the susceptible compartment (Figure~\ref{fig:sir_diagrams} B).
    \item Model without cross-immunity, in which recovery from one variant
    does not prevent later infection by the other variant (Figure~\ref{fig:sir_diagrams} C).
    \item Combined model with no complete cross-immunity, partial asymmetric
    protection against secondary infection, and variant-specific waning
    immunity (Figure~\ref{fig:sir_diagrams} D).
\end{enumerate}

\subsection{Numerical implementation}
\label{sec:implementation}

The SIR systems were solved in Python using \texttt{odeint()} from
\texttt{scipy.integrate}. This routine uses the LSODA algorithm, which switches
adaptively between non-stiff and stiff integration methods. The same numerical
methodology was used for the baseline SIR model and for the three extensions.

The population size was fixed at $N=10^7$. Variant~1 was introduced at time
$T_1$ with 1000 infected individuals, and variant~2 was introduced at time
$T_2$ with 100 infected individuals. For fixed-delay scans, only the difference
$T_2-T_1$ matters, so we set $T_1=0$. For the simulations with normal or
lognormal delay distributions, we used $T_1=0$ and $T_2=15$.

For the baseline complete-cross-immunity model with fixed delays, two parameter
regimes were used:
\begin{itemize}
    \item Regime~1:
        $\beta_1=0.3,
        \beta_2=1.0,
        \gamma_1=\gamma_2=0.2.$
    \item Regime~2:
    $
        \beta_1=0.3,
        \beta_2=0.45,
        \gamma_1=\gamma_2=0.2.
    $
\end{itemize}
For reference, the baseline SIR parameters can also be expressed in terms of
variant-specific basic reproduction numbers,
\begin{equation}
    R_{0,i}=\frac{\beta_i}{\gamma_i}.
\end{equation}
Thus, in both baseline regimes variant~1 has
the basic reproduction number $R_{0,1}=0.3/0.2=1.5$. In Regime~1, the variant~2 gives $R_{0,2}=1.0/0.2=5.0$, while in Regime~2,  $R_{0,2}=0.45/0.2=2.25$. The larger value in
Regime~1 should be interpreted as a modeling choice that allows the later
invading variant to generate a visible replacement wave after substantial
susceptible depletion, rather than as a calibrated epidemiological estimate.

The logit-slope estimator is not, however, an estimator of $R_{0,2}/R_{0,1}$
itself. In the SIR model, the quantity directly connected to the change in
variant frequency is the difference between instantaneous per-capita growth
rates,
\begin{equation}
    g_2(t)-g_1(t)
    =
    \left(\beta_2-\beta_1\right)\frac{S(t)}{N}
    -
    \left(\gamma_2-\gamma_1\right).
\end{equation}
Equivalently, with the effective reproduction number
$R_i(t)=R_{0,i}S(t)/N$, one has
\begin{equation}
    g_i(t)=\gamma_i\left(R_i(t)-1\right).
\end{equation}
Therefore, even when the basic reproduction numbers are fixed parameters, the
relative growth rate changes over time as the susceptible pool is depleted. A
fitted logit slope should therefore be interpreted as a time-window summary of
relative growth, not as a direct estimate of a fixed reproduction-number ratio.

For the baseline model with normal and lognormal distributed delays, we used
$    \beta_1=0.3,
    \beta_2=0.35,
    \gamma_1=0.2,
    \gamma_2=\frac{1}{6}$ ($R_{0,1}=1.5$, $R_{0,2}=2.1$).
The parameters for the three extended models are summarized in
Table~\ref{table_sir_parameters}.

\begin{table}[ht]
\centering
\caption{Parameter values used in the extended SIR model variants. Columns
labeled WI/NCI refer to the waning-immunity (Figure~\ref{fig:sir_diagrams} B) and no-cross-immunity models (Figure~\ref{fig:sir_diagrams} C),
which share the listed transmission and recovery parameters. Columns labeled CM
refer to the combined model with partial asymmetric protection and
variant-specific waning immunity (Figure~\ref{fig:sir_diagrams} D). 
A dash indicates that the parameter is not
part of the corresponding model.}
\label{table_sir_parameters}
\begin{tabular}{lcccc}
\hline
& \multicolumn{2}{c}{Regime~1: }
& \multicolumn{2}{c}{Regime~2: } \\
Parameter & TI/NCI & CM & TI/NCI & CM \\
\hline \hline
$\beta_1$   & 0.3 & 0.3  & 0.3 & 0.3 \\
$\beta_2$   & 1.0 & 0.6  & 0.45 & 0.43 \\
$\gamma_1$  & 0.2 & 0.2  & 0.2 & 0.2 \\
$\gamma_2$  & 0.2 & 0.2  & 0.2 & 0.2 \\
$T_2$       & --  & 30   & --  & 80 \\
$\chi_{21}$ & --  & 0.4  & --  & 0.4 \\
$\chi_{12}$ & --  & 0.6  & --  & 0.6 \\
$\omega_1$  & --  & $1/90$ & -- & $1/90$ \\
$\omega_2$  & --  & $1/75$ & -- & $1/75$ \\
$\omega$    & $1/90$ \big/ -- & -- & $1/90$ \big/ -- & -- \\
\hline
\end{tabular}
\end{table}

After solving the ODE system, daily incidence curves were computed from the
infection terms and discretized into daily bins. A fixed fraction of incident
infections was assumed to be observed, with the same observation fraction for
both variants. This fraction cancels when observed variant proportions are
computed.
For fixed delays, all incidence assigned to variant $i$ was shifted by
$d_i$ days:
\begin{equation}
    y_i(t)
    =
    \rho\,\lambda_i(t-d_i).
\end{equation}
For distributed delays, delayed observations were obtained by convolution with a
discretized delay distribution:
\begin{equation}
    y_i(t)
    =
    \rho
    \sum_{d\geq0}
    \lambda_i(t-d)g_i(d),
\end{equation}
where $g_i(d)$ is the probability of observing an infection by variant $i$ after
a delay of $d$ days. The distribution was discretized into daily bins and
renormalized.

For normally distributed delays,
$D_i\sim\mathcal{N}(\mu_i,\sigma_i^2)$,
the distribution was restricted to
$(\mu_i-2\sigma_i,\mu_i+2\sigma_i)$
before discretization and renormalization. Parameter combinations with
substantial probability mass at negative delays were excluded. In particular,
we required $\mu_i\geq2\sigma_i$ for the varied delay distribution.

In case of lognormal delay  $D_i \sim \mathrm{Lognormal}(\mu_i,\sigma_i^2)$, or 
 $\log D_i \sim \mathcal{N}(\mu_i,\sigma_i^2)$, the delay distribution is positive by construction. Therefore no lower  truncation was required. To exclude unrealistically long observation delays, an upper truncation was applied:  $\log D_i \leq \mu_i + 2\sigma_i$. The results for this case are reported in Supplementary Material.

The delayed observed incidence curves were used to compute
\begin{equation}
    p_2^{\mathrm{obs}}(t)
    =
    \frac{y_2(t)}{y_1(t)+y_2(t)}.
\end{equation}
We then fitted
\begin{equation}
    \operatorname{logit}p_2^{\mathrm{obs}}(t)
    =
    \alpha+s_{\mathrm{obs}}t
\end{equation}
on the interval, in which $p_2^{\mathrm{obs}} \in (0.05,0.95)$.
The same fitting rule was applied to the corresponding undelayed incidence
curves, giving slope $s$. The relative slope error was defined as
\begin{equation}
    \Delta s
    =
    \frac{s_{\mathrm{obs}}-s}{s}.
    \label{eq:relative_slope_error}
\end{equation}
with positive values indicating that delayed observations overestimate the undelayed logit slope, and negative values indicating the opposite.

\section{Numerical results for SIR models}
\label{sec:numerical_results}

The numerical experiments use SIR-type models as controlled examples of
two-population evolutionary dynamics. Variant~1 is introduced first, and
variant~2 is introduced later. The goal is  
to test whether delayed observation can create a mismatch
between the logit slope inferred from observed data and the corresponding slope
from undelayed incidence.

\subsection{Fixed variant-specific delays in the baseline SIR model}
\label{sec:SIR_fixdelay}

In the exponential model, delay changes the intercept of the
log-odds trajectory but not its slope. This does not need to hold in SIR dynamics,
because the susceptible pool changes over time. 
We scanned fixed delay differences $d_2-d_1$ and introduction-time differences
$T_2-T_1$ in the two parameter regimes described above. 

\begin{figure}
\centering
\includegraphics[width=0.9\textwidth]{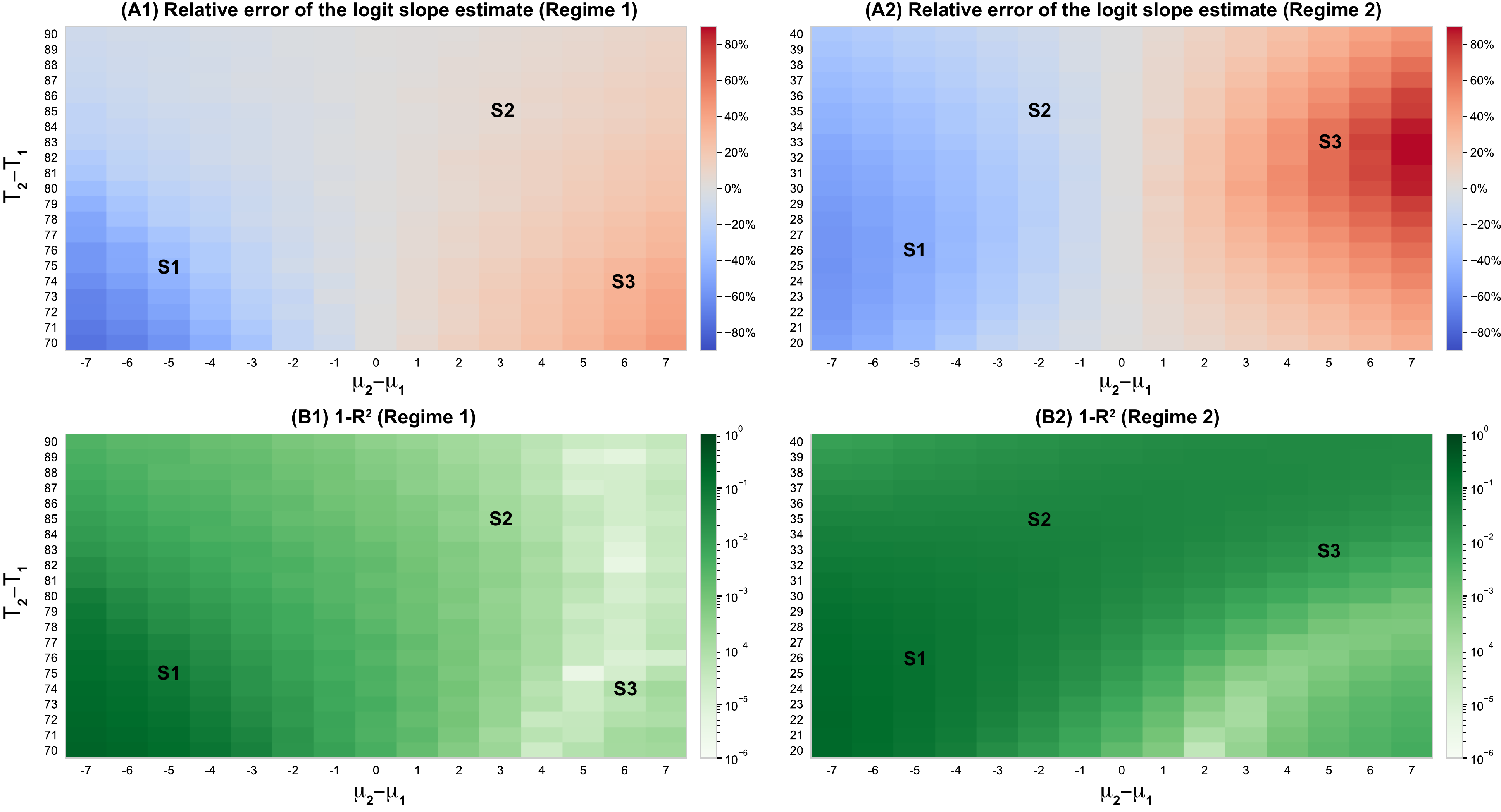}
\includegraphics[width=0.9\textwidth]{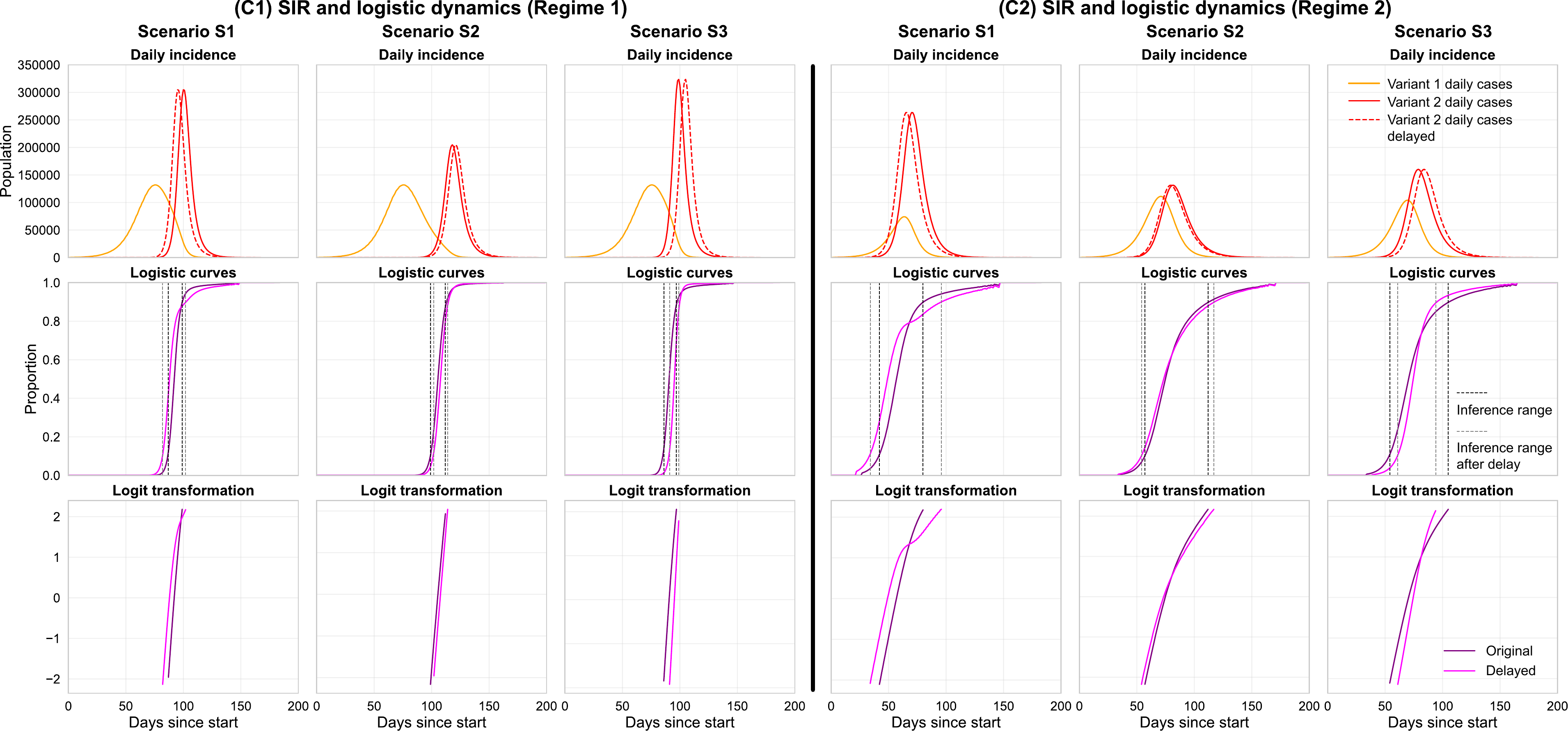}
\caption{Baseline two-variant SIR model with complete cross-immunity and fixed
variant-specific observation delay.
{(A1)} Relative slope error in Regime~1 as a function of the delay difference
$d_2-d_1$ and introduction-time difference $T_2-T_1$.
{(B1)} Corresponding heat map of $1-R^2$, where $R^2$ is computed from the
linear fit to the delayed logit-transformed trajectory.
{(A2)} Relative slope error in Regime~2.
{(B2)} Corresponding heat map of $1-R^2$.
{(C1,C2)} Representative scenarios selected from panels (A1) and (A2). For
each regime, the panels show daily incidence, the delayed proportion of
variant~2, and the corresponding logit transformation. Scenario S1 represents a
large negative slope error, S2 a small slope error, and S3 a large positive
slope error. The numerical values are reported in
Table~\ref{tab:sirbaseline}.}
\label{fig:sirbaseline}
\end{figure}

Figure~\ref{fig:sirbaseline} shows that the fitted logit slope can differ
substantially from the undelayed value. In these simulations, the sign of the
slope error is largely controlled by the delay difference. When variant~1 has
the longer delay, $d_2-d_1<0$, the delayed data tend to underestimate the
undelayed slope. When variant~2 has the longer delay, $d_2-d_1>0$, the delayed
data tend to overestimate it. The magnitude of the error generally decreases
when the introduction-time difference is larger, because the two epidemic waves
overlap less strongly.

The $1-R^2$ panels distinguish two situations. In some parameter regions the
logit fit remains very good, so the bias may not be apparent from goodness of
fit alone. In other regions the delayed replacement trajectory is not well
approximated by a straight line on the logit scale, and the fitted slope has
limited interpretation as a single relative growth advantage.

\begin{table}[ht]
\centering
\caption{Baseline SIR model with complete cross-immunity and fixed
variant-specific delay. Relative slope errors compare delayed and undelayed
logit-slope estimates. The value of $R^2$ measures linearity of the delayed
logit-transformed replacement trajectory. Scenarios S1--S3 are marked in
Figure~\ref{fig:sirbaseline}.}
\label{tab:sirbaseline}
\begin{tabular}{lcc}
\hline
Scenario & Relative slope error & $R^2$ \\
\hline \hline
S1, Regime~1 & $-38.3\%$ & 0.9254 \\
S2, Regime~1 & $5.5\%$   & 0.9998 \\
S3, Regime~1 & $31.1\%$  & 0.9999 \\
S1, Regime~2 & $-47.3\%$ & 0.8812 \\
S2, Regime~2 & $-15.7\%$ & 0.9542 \\
S3, Regime~2 & $60.7\%$  & 0.9841 \\
\hline
\end{tabular}
\end{table}

The key cautionary case is Regime~1, scenario S3. In this example, the relative
slope error is approximately $31\%$, while the delayed logit trajectory remains
almost perfectly linear, with $R^2=0.9999$. This error would therefore not be
detected by inspecting the linearity of the logit-transformed replacement curve
alone. A high-quality logistic fit can does not contradict a biased estimate
 of the logit slope from observed data.

\subsection{Fixed delays in extended SIR models}
\label{sec:SIR_extended_results}

We next tested whether the same qualitative issue appears in the extended SIR
models. These extensions change the immunity structure, and therefore they also
change the shape of the replacement trajectory. Our analysis does not aim to
reproduce specific biological processes quantitatively but is qualitative in nature.

\begin{figure}
\centering
\includegraphics[width=0.98\textwidth]{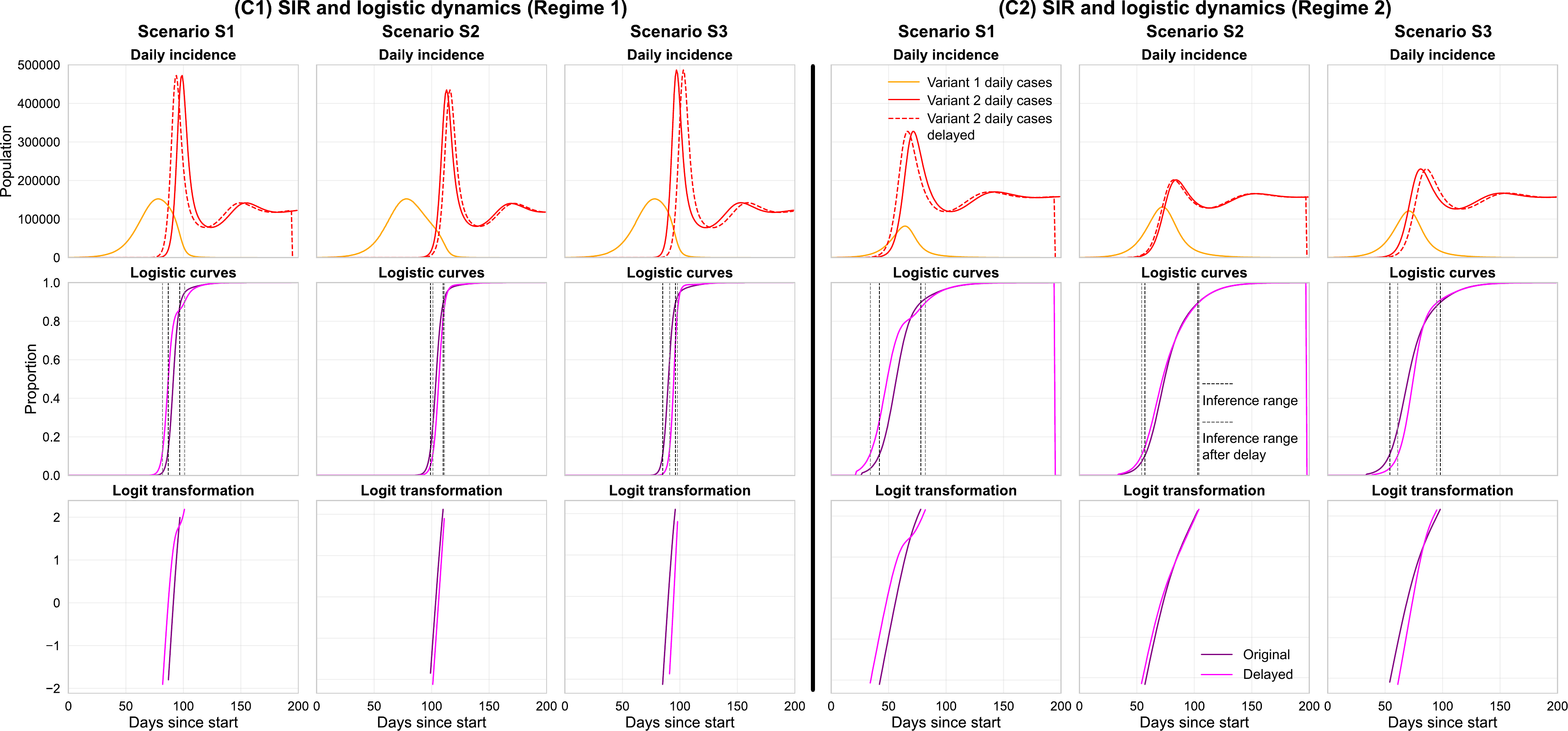}
\caption{Selected scenarios for the temporary-immunity SIR model. For each
scenario, the pair $(d_2-d_1,\;T_2-T_1)$ is reported.
{(C1)} Regime~1: S1 $=(-5,75)$, S2 $=(3,85)$, S3 $=(6,74)$.
{(C2)} Regime~2: S1 $=(-5,26)$, S2 $=(-2,35)$, S3 $=(5,33)$.}
\label{fig:waningappendix}
\end{figure}

\paragraph{Temporary immunity.}
In the temporary-immunity model, see Figure~\ref{fig:sir_diagrams} (B), recovered individuals return to the susceptible
compartment at rate $\omega$. The selected scenarios, summarized in Table~\ref{fig:waningappendix}, show the same qualitative
behavior as the baseline model: delayed observation can change the fitted logit
slope, and some biased trajectories remain close to linear after logit transformation.

\begin{table}[ht]
\centering
\caption{Temporary-immunity SIR model. Relative slope errors compare delayed
and undelayed logit-slope estimates. The table also reports the $R^2$ value of
the linear fit to the delayed logit-transformed trajectory.}
\label{table_sir_wi}
\begin{tabular}{lcc}
\hline
Scenario & Relative slope error & $R^2$ \\
\hline \hline
S1, Regime~1 & $-43.0\%$ & 0.9050 \\
S2, Regime~1 & $8.2\%$   & 0.999994 \\
S3, Regime~1 & $33.2\%$  & 0.9998 \\
S1, Regime~2 & $-37.9\%$ & 0.9123 \\
S2, Regime~2 & $-11.5\%$ & 0.9713 \\
S3, Regime~2 & $47.2\%$  & 0.9827 \\
\hline
\end{tabular}
\end{table}

\paragraph{No cross-immunity.}
In the no-cross-immunity model, see Figure~\ref{fig:sir_diagrams} (C), recovery from one variant does not protect against infection by the other variant. This changes the competition structure.
The Regime 1 scenarios remain close to logit-linear, as shown in Table~\ref{table_sir_nci}, and have small
slope errors. In contrast, several Regime 2 scenarios have very low
$R^2$ values. These cases should not be interpreted as precise bias estimates,
rather, they show that the assumption of a single logit-linear replacement
trajectory can fail.

\begin{figure}
\centering
\includegraphics[width=0.98\textwidth]{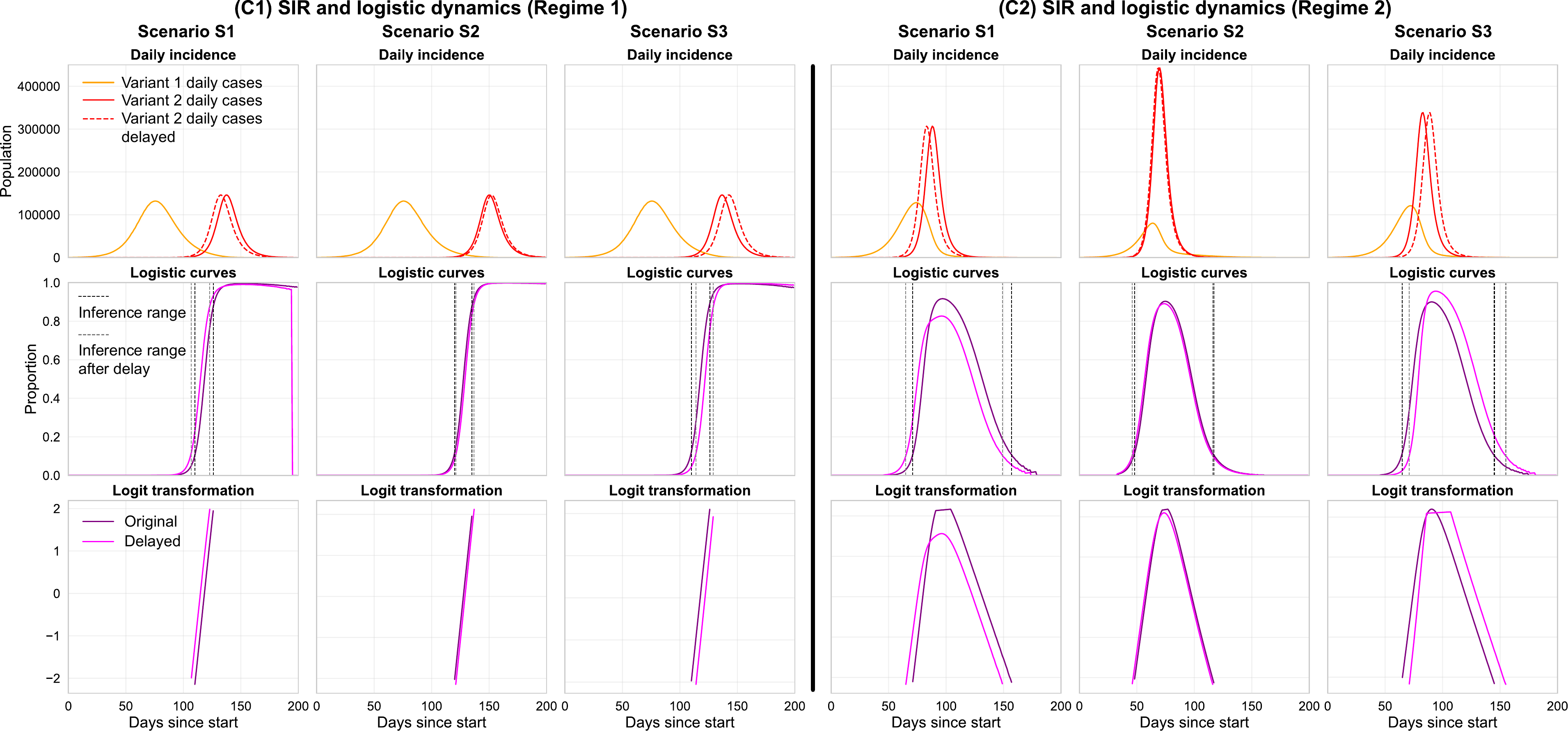}
\caption{Selected scenarios for the no-cross-immunity SIR model. The same
combinations of $(d_2-d_1,\;T_2-T_1)$ as in
Figure~\ref{fig:waningappendix} are shown.}
\label{fig:nocrossappendix}
\end{figure}

\begin{table}
\centering
\caption{No-cross-immunity SIR model. Relative slope errors compare delayed and
undelayed logit-slope estimates. Very low $R^2$ values indicate that the
logit-slope estimator is not a meaningful summary of a single logistic
replacement trajectory.}
\label{table_sir_nci}
\begin{tabular}{lcc}
\hline
Scenario & Relative slope error & $R^2$ \\
\hline \hline
S1, Regime~1 & $-2.7\%$  & 0.9997 \\
S2, Regime~1 & $0.5\%$   & 0.999998 \\
S3, Regime~1 & $3.7\%$   & 0.9999 \\
S1, Regime~2 & $64.2\%$  & 0.0915 \\
S2, Regime~2 & $-1.9\%$  & 0.0746 \\
S3, Regime~2 & $-96.1\%$ & 0.0003 \\
\hline
\end{tabular}
\end{table}

\paragraph{Combined immunity model.}
The combined model, see Figure~\ref{fig:sir_diagrams} (D), includes partial asymmetric protection against secondary
infection together with variant-specific waning immunity. It produces an
intermediate pattern, see Table~\ref{table_sir_cm}. Some Regime 1 scenarios remain close to
logit-linear despite non-negligible slope errors, whereas some
Regime 2 scenarios have weak logit-linearity.

\begin{figure}
\centering
\includegraphics[width=0.98\textwidth]{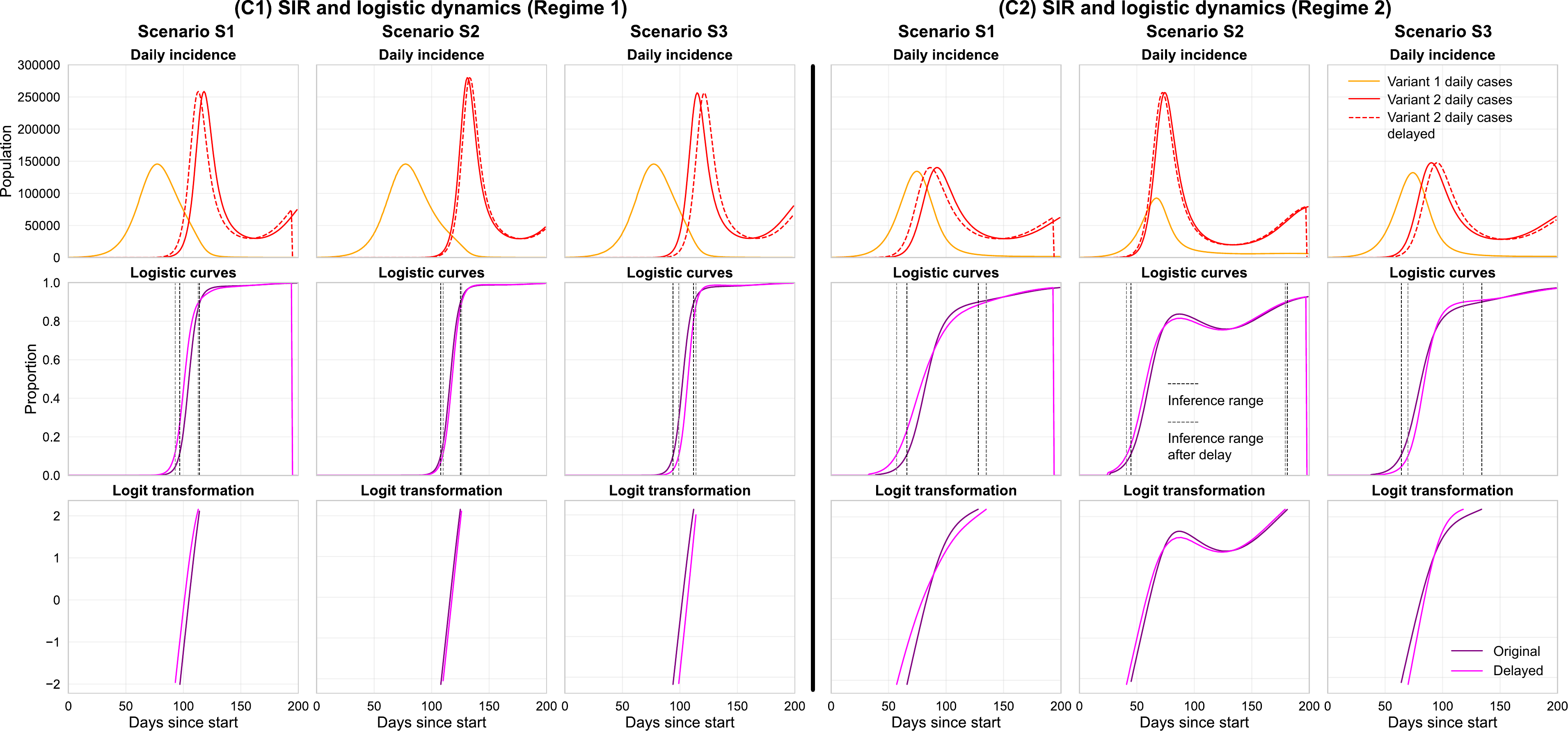}
\caption{Selected scenarios for the combined SIR model with partial asymmetric
protection and variant-specific waning immunity. For each scenario, the pair
$(d_2-d_1,\;T_2-T_1)$ is reported.
{(C1)} Regime~1: S1 $=(-5,82)$, S2 $=(2,96)$, S3 $=(6,81)$.
{(C2)} Regime~2: S1 $=(-3,50)$, S2 $=(-2,33)$, S3 $=(6,48)$.}
\label{fig:combinedappendix}
\end{figure}

\begin{table}
\centering
\caption{Combined SIR model. Relative slope errors compare delayed and
undelayed logit-slope estimates. The table also reports the $R^2$ value of the
linear fit to the delayed logit-transformed trajectory.}
\label{table_sir_cm}
\begin{tabular}{lcc}
\hline
Scenario & Relative slope error & $R^2$ \\
\hline \hline
S1, Regime~1 & $-13.6\%$ & 0.9905 \\
S2, Regime~1 & $2.5\%$   & 0.99996 \\
S3, Regime~1 & $14.7\%$  & 0.9999 \\
S1, Regime~2 & $-22.7\%$ & 0.9453 \\
S2, Regime~2 & $4.8\%$   & 0.4718 \\
S3, Regime~2 & $57.5\%$  & 0.0414 \\
\hline
\end{tabular}
\end{table}

The extended models support the same results as the
baseline model. Delayed observation can either bias an apparently well-fit
logit-slope estimate, or it can lead to replacement trajectories inconsistent with
a logistic function. 

\subsection{Variant-specific distributed delays in the baseline SIR model}
\label{sec:SIR_stochdelay}

The fixed-delay simulations isolate the effect of a difference between the
observation times of the two variants. We next tested whether adding 
spread to the delay distribution changes this conclusion. The main comparison
uses normally distributed delays, while the comparison in Supplementary Material
considers a more realistic lognormal distribution. The case of distributed delays is contrasted with fixed delays
having the same mean. The fixed-delay comparison is defined as 
the zero-variance limit of the normal delay model. Specifically,
$D_i\sim\mathcal{N}(\mu_i,\sigma_i^2)$ converges to a fixed delay
$d_i=\mu_i$ as $\sigma_i\to0$. We therefore represent the fixed-delay case by
$\sigma_i=0$.

Variant~1 was assigned either a distributed delay
$  D_1\sim\mathcal{N}(5,2.5^2) $
or the corresponding fixed delay $d_1=5$. The mean $\mu_2$ and standard
deviation $\sigma_2$ of the delay distribution of variant~2 were then varied,
subject to the truncation and normalization described in
Section~\ref{sec:implementation}.

\begin{figure}
\centering
\includegraphics[width=0.8\textwidth]{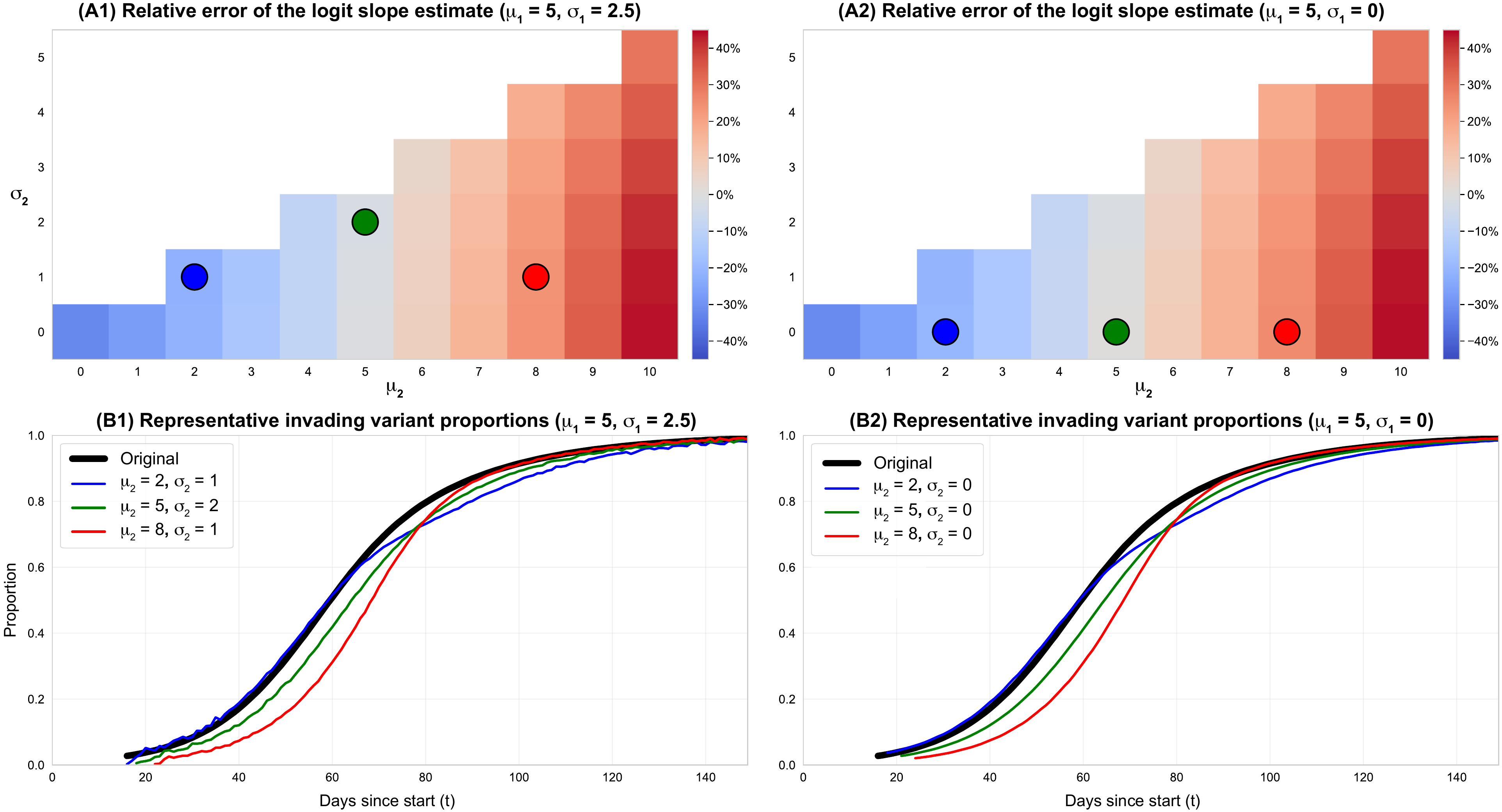}
\caption{Sensitivity of baseline SIR logit-slope estimates to normally
distributed observation delays.
{(A1)} Relative slope error for $D_1\sim\mathcal{N}(5,2.5^2)$ as a function of
the mean $\mu_2$ and standard deviation $\sigma_2$ of
$D_2\sim\mathcal{N}(\mu_2,\sigma_2^2)$.
{(A2)} Corresponding analysis with fixed delay for variant~1,
$d_1=5$. The fixed-delay comparison for variant~2 is represented by
$\sigma_2=0$.
{(B1, B2)} Representative delayed invading-variant proportions for selected
parameter combinations from panels A1 and A2. Only parameter combinations with
$\mu_2\geq2\sigma_2$ are shown, to avoid regimes with substantial probability
mass at negative delays.}
\label{fig:sirnormalnoise}
\end{figure}

\begin{table}
\centering
\caption{Baseline SIR model with normally distributed observation delays. The
table reports logit-slope estimates obtained from delayed synthetic data and
their relative errors with respect to the corresponding undelayed estimate. The
columns labeled random delays use distributed delays for both variants. The
columns labeled fixed delays use fixed delays $d_1=\mu_1$ and
$d_2=\mu_2$.}
\label{tab:sirnormalnoise}
\small
\begin{tabular}{lccccc}
\hline
Variant~1 delay & Variant~2 delay
& \begin{tabular}[c]{@{}c@{}}Delayed slope\\random delays\end{tabular}
& \begin{tabular}[c]{@{}c@{}}Relative error\\random delays\end{tabular}
& \begin{tabular}[c]{@{}c@{}}Delayed slope\\fixed delays\end{tabular}
& \begin{tabular}[c]{@{}c@{}}Relative error\\fixed delays\end{tabular} \\
\hline\hline
$D_1\sim\mathcal{N}(5,2.5^2)$ & $D_2\sim\mathcal{N}(2,1^2)$
& 0.0548 & $-21.5\%$
& 0.0557 & $-20.1\%$ \\
$D_1\sim\mathcal{N}(5,2.5^2)$ & $D_2\sim\mathcal{N}(5,2^2)$
& 0.0684 & $-2.0\%$
& 0.0697 & $0\%$ \\
$D_1\sim\mathcal{N}(5,2.5^2)$ & $D_2\sim\mathcal{N}(8,1^2)$
& 0.0863 & $23.7\%$
& 0.0875 & $25.4\%$ \\
\hline
\end{tabular}
\end{table}

The distributed-delay and fixed-delay comparisons are close in the
selected regimes. When the mean delay of variant~2 is smaller than that of
variant~1, the fitted slope is reduced by about $20\%$ in both cases. When the
mean delay of variant~2 is larger, the fitted slope is increased by about
$25\%$. The variance of the delay distribution changes the exact numerical
value, but in these simulations it does not change the sign or the main
magnitude of the effect. The analogous analysis with lognormally distributed
delays, given in Supplementary Material, leads to the same
qualitative conclusion. The fixed-delay scans should therefore be read
as a transparent reduction of the parameter space, not as an assumption that
real observation delays are fixed.

The numerical examples distinguish two different failures
of logit-slope inference from delayed observations. In the first, the delayed
replacement trajectory remains close to logit-linear, but the fitted slope is
differs from the undelayed event-time slope. This is important for empirical applications, because the usual diagnostic of logit-linearity
does not reveal the bias. In the second, the replacement trajectory is not
approximately logit-linear, and the logistic model is not applicable. 

\section{Discussion}

This work presents a cautionary example for inference from replacement trajectories in evolutionary competition. A common empirical strategy is to fit a linear trend to the logit-transformed frequency of an invading type and to use the fitted slope as a proxy for relative fitness or selective advantage. We show that this procedure can be sensitive to observation delays. If competing populations are observed after type-dependent delays, the apparent logit slope estimated from observed data does not need to match the corresponding slope at the underlying event-time dynamics.

When both types grow exponentially with constant growth rates, and delayed type-specific counts are aggregated before frequencies are formed, the delay changes only the intercept of the log-odds trajectory. The logit slope remains equal to the difference of the exponential growth rates. For time-dependent growth rates, observation delays shift the times at which the growth rates of the two types are evaluated, so type-specific delays can significantly change the relative growth rate inferred from the observed trajectory. Delay-induced bias can appear when the underlying growth rates change over time, but also when boundary effects become important, or when the observation model differs from the aggregate-count delay model considered in this work.

In SIR-type models, the competing variants interact through the susceptible population and the per-capita growth rates vary over time as the epidemic develops. A delay changes which dynamical phase of one variant is compared with which phase of the other. The most important consequence is not only that delayed and undelayed slopes can differ, but that they can differ substantially even when the delayed trajectory is almost perfectly logit-linear. Therefore, a good logistic fit is not, by itself, evidence that the observed-time slope is an unbiased estimate of the underlying event-time relative growth advantage.

In addition, in some regimes, especially when the interaction structure differs from complete cross-immunity or when the invading type arrives early, the replacement trajectory is not well described by a single logistic curve. In these cases, the difference between delayed and undelayed fitted slopes should not be interpreted as a precise bias estimate.

The SARS-CoV-2 analysis illustrates why this distinction matters for observed biological data.
The Alpha$\to$Delta and Delta$\to$Omicron BA.1 replacement waves in the United Kingdom provide clear examples of approximately logistic replacement in observed sequence frequencies. However, GISAID sequence records are observations of collected and submitted samples, not infection events. The empirical logit-slopes therefore describe sequence frequency dynamics on the observation scale. 

The distinction between event time and observation time is also central to epidemiological nowcasting.
Nowcasting methods reconstruct event-time incidence from delayed and incomplete observations.
Although this approach is closely related to ours, it addresses a different inference problem: our focus is on how type-specific observation delays affect relative-frequency inference. In principle, a correctly specified variant-specific nowcasting model could reduce or remove the bias studied here by reconstructing infection-time incidence separately for each variant before estimating the logit slope. However, this requires information about variant-specific delay distributions, sampling probabilities, and reporting processes. Standard nowcasting of total incidence, or nowcasting models that assume a common delay distribution across variants, do not directly address the bias mechanism considered in this work. 

The present analysis has several limitations.
The analysis does not estimate real infection-to-observation delay distributions, and the SIR simulations are intended to provide qualitative rather than calibrated epidemiological results.
The SIR simulations are qualitative examples designed to isolate the mechanism by which observation delay affects apparent selective advantage.
Also, the observation model assumes a
constant observed fraction that is the same for both types. Real surveillance systems may have time-varying and type-dependent sampling probabilities, reporting practices, and data availability. The delay distributions used in the simulations are stylized, and the empirical estimator is intentionally simple: ordinary least-squares regression on logit-transformed frequencies. These choices suffice to demonstrate the mechanism, whereas a calibrated inference framework would require combining explicit delay distributions with likelihood-based models for type-specific counts and quantifying how uncertainty in delays propagates into estimates of relative fitness.

Observation delays can therefore affect inference of relative advantage whenever competing populations have time-dependent growth rates. In particular, a nearly linear logit trajectory does not guarantee that the observed-time slope accurately represents the underlying event-time relative growth advantage. Sensitivity to type-specific observation delays should therefore be considered when logit slopes are used as proxies for relative fitness.

\section*{Declaration of generative AI and AI-assisted technologies in the manuscript preparation process}
During the preparation of this work the authors used ChatGPT (OpenAI) to assist with language editing and improvement of the presentation. After using this tool, the authors reviewed and edited the content as needed and take full responsibility for the content of the published article.

\section*{Acknowledgements}
Part of this research is based on data from GISAID. We gratefully acknowledge all data contributors, i.e., the authors and their originating laboratories responsible for obtaining the specimens, and their submitting laboratories for generating the genetic sequence and metadata and sharing via the GISAID Initiative.  Funded by the EU NextGenerationEU through the Recovery and Resilience Plan for Slovakia under the project No. 09I03-03-V04-00037.

\end{document}